\documentclass[twocolumn,nofootinbib]{revtex4}
\usepackage{graphicx}
\usepackage{latexsym}
\def\be{\begin{equation}}
\def\ee{\end{equation}}
\def\bea{\begin{eqnarray}}
\def\eea{\end{eqnarray}}

\begin{document}
\title{K-essential Leptogenesis}
\author{Mingzhe Li}
  \email{limz@mail.ihep.ac.cn}
\author{Xinmin Zhang}
  \email{xmzhang@mail.ihep.ac.cn}
\affiliation{Institute of High Energy Physics, Chinese
Academy of Sciences, P.O. Box 918-4, Beijing 100039, People's Republic of China}

\begin{abstract}
K-essence is a possible candidate for dark energy of the Universe.
In this paper we consider couplings of k-essence to the matter
fields of the standard electroweak theory and study the effects of
the cosmological CPT violation induced by the CPT violating {\bf
Ether} during the evolution of the k-essence scalar field on the
laboratory experiments and baryogenesis. Our results show that the
matter and antimatter asymmetry can be naturally explained {\it
via} leptogenesis without conflicting with the experimental limits
on CPT violation test. The mechanism for baryogenesis proposed in
this paper provides a unified picture for dark energy and baryon
matter of our Universe and allows an almost degenerate neutrino
mass pattern with a predicted rate on the neutrinoless double beta
decays accessible to the experimental sensitivity in the near future.
\end{abstract}

\maketitle

\hskip 1.6cm
PACS number(s): 98.80.Cq, 11.30.Er
\vskip 0.4cm

There are strong evidences that the Universe is spatially flat and
accelerating at the present time \cite{pel}. The simplest account
of this cosmic acceleration seems to be a remnant small
cosmological constant, however many physicists are attracted by
the idea that a new form of matter, usually called dark
energy \cite{tur} is causing the cosmic accelerating. A simple
candidate for dark energy is a scalar field (or multi scalar
fields), {\it quintessence} \cite{rp,we,fhsw,swz} which includes a
canonical kinetic term and a potential term in the Lagrangian.
Another one is a scalar field with non-canonical
kinetic terms which is called in the literature as
k-essence \cite{pms,coy}. Differing from quintessence, for
k-essence the accelerating expansion of the Universe is driven by
its kinetic rather than potential energy.

Being a dynamical component, the scalar field dark energy is
expected to interact with the ordinary matters.
There are many discussions on the explicit couplings of
quintessence to baryons, dark matter and photons \cite{ca,peccei,amen,peebles},
however as argued in Refs. \cite{ca, peccei} for most of the cases the couplings are strongly
constrained. But there are exceptions. For example,
Carroll \cite{ca} has considered an interaction of form $Q
F_{\mu\nu}{\tilde F^{\mu\nu}}$ with $F_{\mu\nu}$ being the
electromagnetic field strength tensor which has interesting
implication on the rotation of the plane of polarization of light
coming from distant sources. In addition, if the interaction is not
universal, as argued in Ref. \cite{damo}, a sizable coupling is
possible. And the
authors of Ref. \cite{uzan} have also studied the quintessence
non-minimally coupled to gravity.
Recent data on the possible variation of the electromagnetic fine structure
constant reported in \cite{fine}
has triggered interests in studies related to the interactions
between quintessence and the matter
fields.

In a recent paper \cite{li} \footnote{For recent studies , see \cite{trodden}} we introduced a type of interaction
between quintessence and the ordinary matters, then studied its implication in the generation of the
baryon number asymmetry of the universe.
Specifically, we have considered a derivative coupling of the quintessence scalar $Q$ to the matter fields,
\begin{eqnarray}\label{lagr}
{\cal L}_{int}=\frac{c}{M }{\partial_{\mu} Q}J^{\mu}~,
\end{eqnarray}
where $M$ is a cut-off scale which, for example could be the scale of Planck
or Grand Unified Theory (GUT), $c$ is a coupling constant
which characterizes the strength of the quintessence interaction
with the ordinary matter fields in the Standard Model of
the electroweak theory. $J^{\mu}$ is the current of the matter fields,
which in Ref. \cite{li} we take to be the baryon current or the current of
the baryon number minus lepton number for the purposes of baryogenesis and leptogenesis.
The Lagrangian (\ref{lagr}) involves a derivative and
obeys the symmetry $\phi \rightarrow \phi + {\it constant}$ \cite{ca}, so
it will not change the quintessence potential by the quantum corrections.
As shown explicitly in Ref. \cite{li}, with a modified exponential quintessence potential given in
Ref. \cite{as}, the ratio of the baryon number to entropy is given by
\begin{eqnarray}\label{bnumber2}
{n_{B}\over s}|_{T_D}\sim 0.01  c \frac{T_D}{M }~,
\end{eqnarray}
where $T_D$ denotes the epoch when the B-violating interactions freeze
out.

One silent feature of this scenario for baryogenesis is
that the present accelerating expansion and the generation of the matter and antimatter
asymmetry of our universe
is described in a unified way.
Furthermore in this scenario the baryon number asymmetry is generated in
thermal equilibrium \cite{ck} which violates one of the conditions by
Sakharov \cite{sak}. This is due to the existence of the CPT violating {\it Ether}
during the evolution of the quintessence scalar field.

One may wonder if this type of CPT violation will affect the laboratory experiments.
At present time the quintessence field is slowly rolling and $\dot
Q$ is bounded from above. To get the maximal value,
$\dot Q_c$, note that
$\frac{1}{2} {\dot
Q}^2 \leq \rho_Q \leq \rho_c \sim 10^{-47} ~[ {\rm GeV}]^4$. So we have
$\dot Q_c \leq 10^{-23} ~[ {\rm GeV} ]^2$.
The experiment of CPT test with a spin-polarized torsion pendulum \cite{cpt} puts strong limits on
the axial vector background $b_\mu$ which
is defined by ${\cal L}=b_{\mu}{\bar e} \gamma^\mu \gamma_5 e$
\cite{colladay}:
\be
 |{\vec b}| \leq 10^{-28} ~{\rm GeV}~.
\ee
For the time component $b_0$, the bound is relaxed to be at the level of $10^{-25}$
GeV \cite{mp}. Taking the current $J^\mu$ in Eq. (\ref{lagr}) to be
${\bar e} \gamma^\mu \gamma_5 e$, $b_0$ here corresponds to $c \frac{\dot Q}{M}$ and
it requires that $b_0\sim c ~ 10^{-23} \frac{ {\rm GeV}^2}{M}\leq 10^{-25}~{\rm GeV}$.
This puts a constraint on the cutoff scale $M$,
however, if taking $M$ to be around the Planck or GUT scale the CPT
violating effects at the present time is much below the current
experimental sensitivity.

In this paper, we will investigate the possibilities of CPT violation and
its implications in baryogenesis with k-essence field $\phi$. We will firstly show that
the current experimental
limits on the CPT violation rule out the
interaction in Eq. (\ref{lagr}) for k-essence. Then we propose one type of coupling
(see Eq. (\ref{intlagr})) and show that the baryon number asymmetry can be explained naturally {\it via} leptogenesis.
We also study the neutrino mass limits imposed by the leptogenesis of our
model.

We start with a brief review on the properties of
k-essence. Generally we consider the Lagrangian of k-essence:
\be\label{lagr1}
{\cal{L}}_{0}=p(\phi, X)=\frac{1}{\phi^2}\tilde{p}(X)~,
\ee
where
\be\label{xdef}
X={1\over 2}\partial_{\mu}\phi\partial^{\mu}\phi~.
\ee
The pressure of the k-essence field is given by $p(\phi, X)$. The energy
momentum tensor
has the form of the perfect fluid:
\be\label{tmn}
T^{\mu}_{\nu}=(\rho+p)U^{\mu}U_{\nu}-p\delta^{\mu}_{\nu}~,
\ee
where the energy density and 4-velocity are,
\bea
\rho &=&2Xp_{,X}-p=\frac{1}{\phi^2}(2X\tilde{p}_{,X}(X)-\tilde{p}(X))\nonumber\\
&\equiv&\frac{1}{\phi^2}\tilde{\rho}(X)~,\\
U_{\nu} &\equiv& \frac{\partial_{\nu}\phi}{\sqrt{2X}}~,
\eea
where $p_{,X}$ represents the derivative of $p(\phi, X)$ with respect to
$X$.
We will not go into the details on the k-essence model,
rather simply point out
a notable feature of k-essence, $~{\rm i.e.}~$, the attractor-like
behavior \cite{pms,chiba}.
In the attractor regimes $X\simeq const.~$. For example, in the radiation
dominated era, one has\cite{pms},
\bea\label{track}
w(X)&=&\frac{\tilde{p}(X)}{\tilde{\rho}(X)}={1\over 3}~,\nonumber\\
X&=& {1\over 2}\dot\phi^2=const.~.
\eea
Furthermore, the stability of the attractor solution requires \cite{pms}:
\be\label{sound}
c^2_s\equiv \frac{\tilde{p}_{,X}}{\tilde{\rho}_{,X}}>{1\over 3}~.
\ee

For the purpose of baryogenesis, we consider the k-essence field coupled to the matter in the following
way:
\be\label{genint}
{\cal L}_{int}=f(\phi)\partial_{\mu}\phi J^{\mu}~,
\ee
where $f(\phi)$ is a function of $\phi$. Taking $f(\phi)=c/M$, the
interaction above is identical to Eq. (\ref{lagr}) for quintessence. And to study
baryogenesis, we take $J^{\mu}$ to be baryon current,
then we calculate $\dot \phi$ and the baryon number asymmetry for a given specific model.
For example, we consider the model proposed in Ref. \cite{pms},
\bea\label{example}
\tilde{p}(X)&=&M_p^6[-2.01+2\sqrt{1+\frac{X}{M_p^4}}+3\times
10^{-17}(\frac{X}{M_p^4})^3\nonumber\\
&~&-10^{-24}(\frac{X}{M_p^4})^4]~,
\eea
where $M_p$ is the reduced Planck mass ($M_p^2\equiv \frac{3m_{pl}^2}{8\pi}$,
and the Planck mass $m_{pl}\equiv\frac{1}{\sqrt G}=1.22\times 10^{19}$ GeV).
In the radiation dominated era, the positive real solutions to Eq.
(\ref{track}) are $X/M_p^4=0.02,~9.3\times 10^6,~1.2\times 10^7$.
When considering the stability condition (\ref{sound}) and
the requirement that the energy density of k-essence is smaller than the critical density of the Universe
$\Omega_{\phi}<1$ (the representation of $\Omega_{\phi}$ in the radiation epoch is derived blow,
see Eq. (\ref{reassume2})),
the solution $X/M_p^4=1.2\times 10^7$ is obtained. Using $X={1\over
2}\dot\phi^2$, we have
\bea\label{rad}
\dot \phi \sim  580m_{pl}^2~.
\eea
And the ratio of the baryon number density to entropy density is given by
\be\label{nb}
\frac{n_B}{s} \sim 0.01~c \frac{\dot\phi}{M T}~.
\ee
Since $\dot\phi$ is as large as $580 m_{pl}^2$, it is quite easy to have
$n_B/s \sim 10^{-10}$ for both $M$ and $T$ in the range of 100
GeV to the Planck scale.
However we will show below that the
large value of $\dot\phi$ at the present time induces CPT violation at the
level which has conflicted already with the experimental limits.

At the present time $\dot\phi$ is
smaller than the value in the radiation epoch given in Eq. (\ref{rad}), however, it is still quite
big. Given the specific model in Eq. (\ref{example}), the authors
of Ref. \cite{pms} have numerically studied the evolution of the
equation of state $w(z)$, and obtained that at the present epoch
$w\simeq -0.77$. Using the relation
$w=\frac{\tilde{p}(X)}{\tilde{\rho}(X)}$,
we can evaluate that $\dot\phi \sim 0.006 m_{pl}^2$.
As argued above in the fifth paragraph of this paper, the non-zero value of $\dot\phi$ will induce CPT
violating effect in the electron system if the k-essence field coupled to
the electron current
$J^{\mu}_e=\bar{e}\gamma^{\mu}\gamma_5e$ through the same way as
$\frac{c}{M}\partial_{\mu}\phi J^{\mu}_e$.
Since $\dot\phi\sim 0.006 m_{pl}^2$, even taking the cut-off scale $M$ to be the Planck mass scale,
we get $b_0=c\dot \phi/m_{pl}$ which is as large as order of
$0.7~c\times 10^{17}$ GeV and will be $10^{42}$ times bigger than the current experimental limit \cite{mp}
unless $c$ is fine tuned to be smaller than the order of $10^{-42}$. With
such a small $c\sim 10^{-42}$, the baryon number asymmetry
generated by Eq. (\ref{nb}) will be much smaller than the
observational data. In the arguments above
we have assumed that the coefficients $c$ are equal in magnitude for
k-essence couplings to the baryon as well as the electron current.
This assumption is reasonable and natural in the sense of without
fine tuning.

Before considering a different possibility of the function $f(\phi)$, we note that in
the small-$X$ regime, the Lagragian in Eq. (\ref{lagr1}) can be transformed to a Lagrangian
of a scalar field with a canonical kinetic energy and an exponential potential.
To the leading order of $X / M^4$, where $M$ is the Planck scale $m_{pl}$ (or the reduced
Planck scale $M_p$ in Eq. (\ref{example})), $\tilde{p}(X)$ can be expanded in general as
\bea
\tilde{p}(X)=aM^6+bM^6 \frac{X}{M^4}~,
\eea
where $a,b$ are dimensionless parameters. For the model we quoted above in Eq. (\ref{example}),
$a=-0.01 $ and $b = 1$. As long as $b>0$, Eq. (\ref{lagr1}) can be changed to
\be\label{quin}
{\cal{L}}_{0}={1\over 2} \partial_{\mu}\psi\partial^{\mu}\psi-
V_{0} \exp{[-\frac{2\psi}{\sqrt{b}M}]}~,
\ee
with the help of the field redefinition given below:
\bea\label{tran}
\partial_{\mu} \psi
&=&\sqrt{b}\frac{M}{\phi}\partial_{\mu}\phi\nonumber\\
\psi &=&\sqrt{b}M\int d\ln \phi~.
\eea
Since ${\cal L}_0$ in Eq. (\ref{quin}) describes a scalar field $\psi$ with an canonical kinetic term,
one expects the form of the scalar $\psi$ coupled to the matter field to
be:
\be\label{intlagr1}
{\cal L}_{int}=c\frac{\partial_{\mu}\psi}{M}J^{\mu}~.
\ee
Given Eq. (\ref{tran}), in terms of the scalar $\phi$, Eq.
(\ref{intlagr1}) becomes
\be\label{intlagr}
{\cal L}_{int}=\frac{c'}{\phi}{\partial_{\mu}\phi} J^{\mu}~,
\ee
which corresponds to
\be\label{fphi}
f(\phi)=\frac{c\sqrt{b}}{\phi}\equiv \frac{c'}{\phi}~.
\ee
The arguments above show that if we choose $f(\phi)$ in the form
given in Eq. (\ref{fphi}), ${\cal L}_{int}$ in Eq. (\ref{intlagr})
will coincide with Eq. (\ref{lagr}) in the limit of small $X$ when the kinetic term normalized canonically.
We note that in k-essence models, $X$ is actually large. Since it
remains a question how to derive a realistic k-essence model (for example, the one given in (\ref{example}))
from a fundamental physics, we consider the coupling (\ref{intlagr})
for a study only in the sense of phenomenology.
In the following we will focus on the phenomenological implication of
${\cal L}_{int}$ in Eq. (\ref{intlagr}) in baryogenesis. Our
results show that with $f(\phi)$ in Eq. (\ref{fphi}), the current experimental limits on CPT test
can be satisfied and the enough baryon number asymmetry can be generated. For the simplicity
of notation, we will drop the prime of $c'$ in the rest of the
paper.

Taking $J^{\mu}=J^{\mu}_B$, during the evolution of the spatial
flat Friedmann-Robertson-Walker Universe, ${\cal L}_{int}$ in Eq. (\ref{intlagr})
generates an effective chemical potential $\mu_{b}$ for baryons:
\bea\label{mub}
\frac{c}{\phi}{\partial_{\mu}\phi} ~J^{\mu}\rightarrow
c\frac{\dot\phi}{\phi}n_{B}=c\frac{\dot\phi}{\phi}(n_{b}-n_{\bar{b}})~,\nonumber\\
\mu_{b}=c\frac{\dot\phi}{\phi}=-\mu_{\bar{b}}~.
\eea
In thermal equilibrium, the net baryon number doesn't vanish as long as $\mu_b\neq 0$ (when $T\gg
m_{b}$) \cite{kt}:
\be\label{nb1}
n_{B}=\frac{g_{b}T^3}{6\pi^2}[\pi^2(\frac{\mu_{b}}{T})+(\frac{\mu_{b}}{T})^3]~,
\ee
 where $g_b$ is the number of intrinsic degrees of freedom of
baryon. The final ratio of the baryon number to entropy is
\bea\label{nb2}
\frac{n_{B}}{s}|_{T_D}\simeq\frac{15g_b}{4\pi^2g_{\star}}\frac{c
\dot\phi}{\phi T_D}~,
\eea
where the cosmic entropy density is $s=\frac{2\pi^2}{45}g_{\star}T^3$ and $g_{\star}$ counts
the total degrees of freedom of the relativistic particles in the Universe at $T_D$
when the B-violating interactions freeze out.

The value of the effective chemical potential can be obtained by solving the
equation of motion for k-essence field which in general is given by:
\bea\label{eqm}
& &(\tilde{p}_{,X}+2X\tilde{p}_{,XX})\ddot\phi+3H\tilde{p}_{,X}
\dot\phi-\frac{2}{\phi}(2X\tilde{p}_{,X}-\tilde{p})=\nonumber\\
& &-c\phi(\dot n_{B}+3Hn_{B})~,
\eea
where $\tilde{p}_{,XX}$ represent the second
derivative of $\tilde{p}(X)$ with respect to $X$. Given
Eqs. (\ref{mub}), (\ref{nb1}) and the Hubble constant during the radiation dominated
epoch,
\be\label{hubble}
H=\frac{1}{2t}=1.66g_{\star}^{1/2}\frac{T^2}{m_{pl}}~,
\ee
the right-hand side of Eq. (\ref{eqm}) can be rewritten as
\be
-c\phi(\dot
n_{B}+3Hn_{B})=-\frac{c^2g_{b}T^2}{6}(\ddot\phi+H\dot\phi-{2X\over
\phi})~.
\ee
So if $T^2$ is much smaller than $\tilde{p}_{,X}$ we can drop off the right-hand side of Eq. (\ref{eqm}).
This means that under this condition the interaction between baryons and
k-essence will not change the main properties of k-essence described in
Ref. \cite{pms}.
Since k-essence has the feature of attractor behaviour shown in Eq. (\ref{track}),
we obtain from Eq. (\ref{eqm}) that
\bea
\frac{\dot\phi}{\phi}=2H=3.32g_{\star}^{1/2}\frac{T^2}{m_{pl}}~,
\label{reassume1}
\eea
and the fraction of the energy density of k-essence is
\bea
\Omega_{\phi}&\equiv &\frac{8\pi\rho(\phi, X)}{3H^2m_{pl}^2}=\frac{8\pi\tilde{p}}
{H^2m_{pl}^2\phi^2}\nonumber\\
&=& \frac{8\pi\tilde{p}_{,X}}{m_{pl}^2}~.\label{reassume2}
\eea
The constraint on $\Omega_{\phi}$ by BBN \cite{bhm} is $\Omega_{\phi}<0.045$.
Taking $\Omega_{\phi}\sim 10^{-2}$,
we have $\frac{\tilde{p}_{,X}}{m_{pl}^2}\sim 10^{-4}-10^{-3}$. In
obtaining the equations above, we have assumed that
$\frac{T^2}{\tilde{p}_{,X}}\ll 1$. To verify this inequality, note
that
\be
\frac{T^2}{\tilde{p}_{,X}}=\frac{T^2}{m_{pl}^2}\frac{m_{pl}^2}{\tilde{p}_{,X}}\sim
\frac{T^2}{m_{pl}^2}(10^3-10^4)~,
\ee
and one can see that $T^2$ will be less than $\tilde{p}_{,X}$ for $T<10^{-2} m_{pl}$.
Moreover, for the temperature $T$ in the range which we are
interested in for Baryogenesis, $T\sim 10^{10}$ GeV (see Eq. (\ref{result}) below),
$T$ is much smaller than $\tilde{p}_{,X}$
and we can safely neglecte the right-hand side in solving Eq. (\ref{eqm}).

With the value of the effective chemical potential, we arrive
at a final expression of the baryon number asymmetry:
\bea
\frac{n_{B}}{s}|_{T_D}&=&\frac{15}{2\pi^2}\frac{c g_{b}H(T_D)}{g_{\star}T_D}\nonumber\\
&\simeq &1.26 c g_{b} g_{\star}^{-1/2} \frac{T_D}{m_{pl}}\sim
0.1c\frac{T_D}{m_{pl}}~.
\eea
In the calculations above, we have used $g_b\sim {\cal O}(1)$ and $g_{\star}\sim {\cal O}(100)$.
Taking $c\sim {\cal O}(1)$, $n_B / s\sim 10^{-10}$ requires the decoupling temperature $T_D$
to be in the order of:
\be\label{result}
T_D\sim 10^{-9} m_{pl}\sim 10^{10}~ \rm{GeV}~.
\ee
A value of $T_D$ at or larger than $10^{10}$ GeV can be achieved in GUT easily.
However, if the B-violating interactions conserve $B-L$, the asymmetry
generated will be erased by the
electroweak Sphaleron \cite{manton}. In this case $T_D$ is as low as around 100 GeV
and $n_B / s$ generated will be of the order of $10^{-18}$.
So now we turn to leptogenesis \cite{lepto,zhang}.
We take $ J^{\mu}$ in Eq. (\ref{intlagr}) to be $J_{B-L}^\mu$.
Doing the calculations with the same procedure as above for $J^{\mu} =
J^{\mu}_{B}$
we have the final asymmetry of the baryon number minus lepton
number
\begin{eqnarray}\label{fnumber2}
 {n_{B-L}\over s}|_{T_D}\sim 0.1  c \frac{T_D}{M }.
\end{eqnarray}
The asymmetry $n_{B-L}$ in (\ref{fnumber2})  will be converted to baryon
number
asymmetry when electroweak
Sphaleron $B+L$
interaction is in thermal equilibrium which happens for
temperature in the range of $10^2 ~{\rm GeV}
\sim 10^{12}{\rm GeV}$. $T_D$ in (\ref{fnumber2}) is the temperature below
which the $B-L$ interactions freeze out.

Now we study the limits on the neutrino masses in our model. In
the scenario of leptogenesis studied in this paper, the $B-L$ number asymmetry is
generated in the thermal equilibrium. This requires that the rate of $B-L$ violating
interactions be larger than the Hubble expanding rate for the temperature above $T_D$.

In the Standard Model of the electroweak theory, $B-L$ symmetry is exactly
conserved, however many models beyond the standard model, such as
Left-Right symmetric model predict the violation of the $B-L$ symmetry.
In this paper we take an
effective Lagrangian approach and parameterize the $B-L$ violation by higher
dimensional operators. There are many operators which violate $B-L$
symmetry, however at dimension 5 there is only one operator,
\begin{eqnarray}\label{lepvio}
{\cal L}_{\not L} = \frac{2} { f } l_L l_L \chi \chi +{\rm H.c.}~,
\end{eqnarray}
where $f$ is a scale of new physics beyond the Standard Model which
generates the $B-L$ violations, $l_L$
and $\chi$ are the left-handed lepton and Higgs doublets respectively.
When the Higgs
field gets a vacuum expectation value $< \chi > \sim v $, the
left-handed neutrino receives a majorana mass $m_\nu \sim
\frac{v^2}{f}$.

\begin{figure}
\includegraphics[scale=0.3]{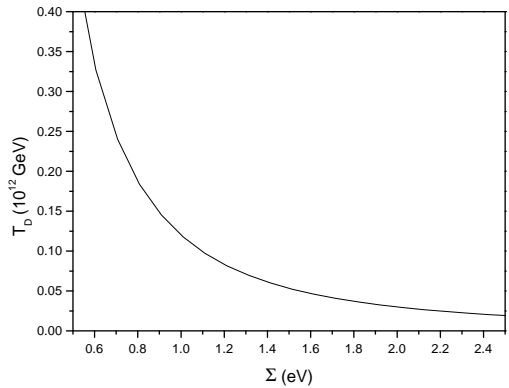}
\caption{\label{fig1} Plot of $T_D$ (in unit of $10^{12}$ GeV) as a function of $\Sigma$ (in unit of eV).}
\end{figure}

In the early universe the lepton number violating rate induced by the
interaction in (\ref{lepvio}) is \cite{sarkar}
\begin{eqnarray}
  \Gamma_{\not L} \sim
    0.04~ \frac{T^3}{ f^2 }~.
\end{eqnarray}
Since $\Gamma_{\not L}$ is proportional to $T^3$, for a given $f$,
namely the neutrino mass, $B-L$ violation will be more efficient at high temperature than at low temperature.
Requiring this rate be larger than the Universe expansion rate
$\sim 1.66 g_{\star}^{1/2}T^2/ m_{pl}$ until the temperature $T_D$,
we obtain a $T_D$-dependent lower limit on the neutrino mass:
\begin{eqnarray}
   \sum_i m_i^2  = {( 0.2 ~{\rm eV} ( { \frac{10^{12}~{\rm GeV}}{T_D} })^{1/2})}^2.
\end{eqnarray}
Taking three neutrino masses to be approximately degenerated, ~{\rm i.e.}~, $m_1 \sim m_2 \sim m_3\sim
{\bar m}$ and defining $\Sigma = 3 {\bar m}$, in Fig. \ref{fig1} we plot
the freezing out temperature $T_D$ as a function of $\Sigma$. One can see
that for $T_D \sim 10^{10}$ GeV, three neutrinos are expected to
have masses $\bar{m}$ around ${\cal O}(1~{\rm eV})$. Numerically, taking
$T_D=1.0 \times 10^{10}$ GeV, we have $\bar{m}=1.2$ eV, for
$T_D=5.0 \times 10^{10}$ GeV, $\bar{m}=0.52$ eV. The
current cosmological limit comes from WMAP \cite{neutrino1}. The analysis of Ref. \cite{neutrino1}
gives $\Sigma<0.69$ eV. Another analysis shows, however that $\Sigma<1.0$ eV \cite{neutrino2}.
These limits on the neutrino masses requires
$T_D$ be larger than $2.5\times 10^{11}$ GeV or $1.2\times 10^{11}$ GeV. The almost
degenerate neutrino masses required by the leptogenesis of this
model will induce a rate of the neutrinoless double beta decays
accessible for the experimental sensitivity in the near future \cite{beta}.
Interestingly, a recent study \cite{allen} on the cosmological data showed a
preference for neutrinos with degenerate masses in this range.

Assuming the k-essence scalar couples to the electron axial current the same as Eq. (\ref{intlagr}),
we can estimate the CPT-violation effect on the laboratory experiments. From the studies of Ref. \cite{pms},
at the present time, our Universe is approaching the k-attractor phase (labeled by the subscript $k$), where
$\Omega_{\phi k}\rightarrow 1$ and $w(X_k)=const. <-\frac{1}{3}$.
So in the near future, one has $\frac{\dot\phi_k}{\phi_k}=\frac{3}{2}(1+w_k)H_k$.
Thus the current value of $\frac{\dot\phi_0}{\phi_0}$ is about $\sim H_0$, and the induced CPT-violating $b_0$ is
\be
b_0\sim c\frac{\dot\phi_0}{\phi_0}\sim c H_0\leq
10^{-42}~\rm{GeV}~,
\ee
which is much below the current experimental limits.

In summary we have studied in this paper the possibility of
baryogenesis in the framework of k-essence as dark energy, and
explicitly showed that the baryon number asymmetry, $n_b / s \sim
10^{-10}$ can be generated via leptogenesis.
Our scenario provides a unified description for the present
accelerating and the generation of the
baryon number asymmetry of our universe by the dynamics of
k-essence.

{\bf{Acknowledgments:}} We would like to thank the innominate referee for comments and suggestions.
This work is supported in part by National Natural
Science Foundation of China and by Ministry of
Science and Technology of China under Grant No. NKBRSF G19990754.

{}

\end{document}